\documentclass[prl,amsmath,twocolumn,showpacs,superscriptaddress]{revtex4}
\usepackage{subfigure}
\newlength{\picwidth}
\usepackage{graphicx}% Include figure files
\usepackage{multirow}
\usepackage{amsmath}
\usepackage{amssymb}
\usepackage{bbold}
\usepackage{color}
\begin{document}

\newcommand{\bra}{\langle}
\newcommand{\ket}{\rangle}
\newcommand{\bs}{\boldsymbol}
\newcommand{\oli}{\overline}
\newcommand{\beq}{\begin{equation}}
\newcommand{\eeq}{\end{equation}}
\newcommand{\bea}{\begin{eqnarray}}
\newcommand{\eea}{\end{eqnarray}}

\title{From closed to open 1D Anderson model: Transport versus spectral statistics}

\author{S. Sorathia}
\affiliation{Instituto de F\'{\i}sica, Universidad Aut\'{o}noma de
Puebla - Apartado Postal J-48, Puebla, Pue., 72570, M\'{e}xico}
\author{F.M. Izrailev}
\email{felix.izrailev@gmail.com}
%\altaffiliation[Also at ]{NSCL and Dept. of Physics and Astronomy, Michigan State %University.}
\affiliation{Instituto de F\'{\i}sica, Universidad Aut\'{o}noma de
Puebla - Apartado Postal J-48, Puebla, Pue., 72570, M\'{e}xico}
\affiliation{NSCL and Dept. of Physics and Astronomy, Michigan State
University - East Lansing, Michigan 48824-1321, USA}
\author{V.G. Zelevinsky}
\affiliation{NSCL and Dept. of Physics and Astronomy, Michigan State
University - East Lansing, Michigan 48824-1321, USA}
\author{G.L. Celardo}
\affiliation{Dipartimento di Matematica e Fisica, Universit\`a Cattolica, via Musei 41, 25121, Brescia, and I.N.F.N., Sezione di Pavia, Italy}
\date{\today}% It is always \today, today,
             %  but any date may be explicitly specified

\begin{abstract}
Using the phenomenological expression for the level spacing distribution with only
one parameter, $0 \leq \beta \leq \infty$, covering all regimes of chaos and complexity in a quantum system, we show that transport properties of the one-dimensional Anderson model of finite size can be expressed in terms of this
parameter. Specifically, we demonstrate a strictly linear relation between $\beta$ and the normalized localization length for the whole transition from strongly localized to extended states. This result allows one to describe all transport properties in the open system entirely in terms of the parameter $\beta$ and strength of coupling to continuum. For non-perfect coupling, our data show a quite unusual interplay between the degree of internal chaos defined by $\beta$, and degree of openness of the model. The results can be experimentally tested in single-mode waveguides with either bulk or surface disorder.
\end{abstract}


%\keywords{Suggested keywords}%Use showkeys class option if keyword
                              %display desired
\maketitle

%%%%%%%%%%%%%%%%%%%%%%%%%%%%%%%%%%%%%%%%%%%%%%%%%%%%%%%%%%%%%%%%%%%%%%%%%%%%%%%%%%%%%%%%%%%%%%%%%%%%%%%%%%%%%%%%%

\section{Introduction} In spite of a remarkable progress in understanding statistical properties of quantum systems, either deterministic or disordered, one of the important problems still awaits for detailed analysis. The specific question is: to what extent one can predict scattering properties of a complex {\it open} system, if we know global properties of eigenstates and energy spectra of the corresponding {\it closed} system? This problem was solved for a specific case of closed systems with maximal chaotic properties described by fully random matrices. A proper mathematical tool in these studies is based on the effective non-Hermitian Hamiltonians of a certain structure \cite{MW69}, and many analytical results have been obtained, see, for example, \cite{supersym,SZ89,fyodorov99}. The key point in this method is that the scattering matrix of an open system is expressed in terms of eigenvalues and eigenfunctions of the related closed system along with their decay amplitudes.

A much more difficult problem emerges if the closed system is not fully chaotic being characterized by %an$
additional parameters related to the degree of chaos. Recently, this problem was analyzed numerically in Refs.~\cite{our,SISZB08,SICZB09} where the global characteristics of scattering or signal transmission through a system have been studied in dependence on two control parameters, the degree of internal chaos and the strength of coupling to continuum. In particular, it was found that, independently of the degree of chaos, the increasing continuum coupling leads the system from the quasi-stable regime of isolated narrow resonances to the ``super-radiant" regime of overlapping resonances coexisting with long-lived compound states. However, the specific features of this evolution may critically depend on the degree of chaos and therefore the observation of the signal transmission provides important information on regular or chaotic character of intrinsic dynamics.

The above studies have been performed with the use of random matrices (canonical Gaussian ensembles or two-body random interactions). Some statistical assumptions may be questionable in application to realistic physical systems. Below we study the 1D Anderson model, paying main attention to the relation between the scattering properties of an open model and those of eigenstates and spectral statistics of the closed model. We discover unexpected effects that give a new insight to the problem of scattering through finite disordered systems.

%%%%%%%%%%%%%%%%%%%%%%%%%%%%%%%%%%%%%%%%%%%%%%%%%%%%%%%%%%%%%%%%%%%%%%%%%%%%%%%%%%%%%%%%%%%%%%%%%%%%%%%%%%%%%%%%%%%%

\section{The model}  
\begin{figure}[h!]
\centerline{\includegraphics[width=1.0\linewidth,angle=0]{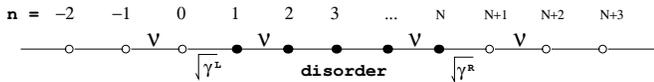}}
\caption{Disordered $1D$ lattice consisting of $N$ sites connected at
both ends to ideal semi-infinite tight-binding leads.}
\label{1Dx}
\end{figure}

The tight-binding Anderson model is often used to describe electron propagation in random media. In the 1D case the corresponding Hamiltonian with diagonal disorder takes the standard form,
\begin{equation}
H_{mn} = \epsilon_n ~\delta_{mn}-\nu
(\delta_{m,n+1}+\delta_{m,n-1}).
\label{Htb}
\end{equation}
Here $\nu$ is the hopping amplitude connecting the nearest sites (in what follows we set $\nu=1$); the site energies $\epsilon_n$ are assumed to be uniformly distributed in the interval $[-W/2,W/2]$ giving rise to the disorder variance $\sigma^2=W^2/12$. Our interest is in the transmission properties through the samples of {\it finite size} $N$ with the arbitrary {\it coupling amplitudes} $\sqrt{\gamma^L}$ and $\sqrt{\gamma^R}$ connecting the left and right edges with attached semi-infinite {\it ideal} leads in which $\epsilon_n=0$, see Fig.~\ref{1Dx}. For zero disorder, open tight-binding models were studied in \cite{SZ92,SR03}. For non-zero disorder, so far, the main interest was related to the statistics and distributions of resonances for one open channel \cite{VZ05,TB}; the relation of the resonances to the transport properties was studied in \cite{celardo10}.

Without disorder, the spectrum of the closed chain consists of Bloch waves with the nodes at the ends and energies inside the band $|E|\leq 2$. In the limit $N \to \infty$ and for weak disorder, $\sigma^2 \ll 1$, all eigenstates are exponentially localized with the characteristic length $l_{\infty}(E)$ given by the Thouless relation \cite{T79},
\beq
l_{\infty}^{-1}(E)=\frac{\sigma^{2}}{8(1-E^2/4)} .
\label{loc}
\eeq
This expression is valid everywhere apart from the vicinity of the band edges, $|E| =2$, and band center, $E=0$.

%%%%%%%%%%%%%%%%%%%%%%%%%%%%%%%%%%%%%%%%%%%%%%%%%%%%%%%%%%%%%%%%%%%%%%%%%%%%%%%%%%%%%%%%%%%%%%%%%%%%%%%%%%%%%%%%%%%%

\section{Level repulsion in a closed model} To quantify the degree of chaos in the finite samples with no continuum couplings, $\gamma^L=\gamma^R=0$, we employ the well known results of the random matrix theory (RMT). In deterministic quantum models with chaotic behavior in the classical counterpart, chaos is usually characterized by the Wigner-Dyson (WD) distribution $P(s)$ of normalized spacings $s$ between the nearest energy levels. In the opposite case of integrable classical counterparts, $P(s)$ is typically close to the Poisson distribution. Therefore, one can take the distribution $P(s)$ as a measure of chaos in the closed model.

For zero disorder, $\epsilon_n=0$, the energy spectrum near the band center is equidistant, so that $P(s)\rightarrow \delta(s-1)$, the eigenstates are extended and regular (standing waves). In the limit of  strong disorder, all eigenstates are effectively localized on the scale of the sample size $N\gg 1$, thus the form of $P(s)$ should be close to the Poisson. In between these limits, for $l_{\infty} \propto N$, it is natural to expect a kind of ``intermediate statistics" which can be compared with the WD-distribution. However, the latter is known to emerge when the eigenstates are ``chaotic" with random fluctuations of their components, as it happens in ensembles of full random matrices \cite{brody81}. It is also known that chaotic eigenstates appear in quasi-dimensional models described by band random matrices, when the localization length is larger than the sample size. Such a situation physically corresponds to the diffusion of wave packets \cite{quasi}. In contrast, in our model described by tri-diagonal matrices, see Eq. (\ref{Htb}), the diffusion scale is absent since the localization length is proportional to the mean free path. For this reason, in the theory of disordered solids an emergence of the WD-distribution for one-dimensional Anderson model of finite size is of special interest \cite{FM94}.

In order to describe  the entire evolution of $P(s)$ as a function of the strength of disorder,
we use the phenomenological expression suggested in Ref.~\cite{CIM91},
\bea
P_{\beta}(s)=B_1 z^{\beta}(1+B_2 \beta z)^{f(\beta)}
\exp\left[-\frac{1}{4}\beta z^2-\left(1-\frac{\beta}{2}\right)z\right],
\label{Psb}
\eea
where
\beq
f(\beta)=\beta^{-1}2^\beta\left(1-\frac{\beta}{2}\right)-0.16874.
\label{f-beta}
\eeq
Here $z=\pi s/2$ and the parameters $B_1$ and $B_2$ are determined by the normalization
conditions,
\beq
\int_0^{\infty}P_{\beta}(s)ds = \int_0^{\infty}sP_{\beta}(s)ds=1.
\label{norm}
\eeq

The above formula (\ref{Psb}) was suggested in Refs.~\cite{CIM91,I93} by using the analytical expressions derived by Dyson \cite{D62} for the classical gas of two-dimensional charged particles moving on a ring at temperature $1/\beta$. This model was found to be very effective since for the values $\beta=1,2,4$ the partition function giving the probability to find the particles at specific positions, coincides with the partition function for eigenvalues in the canonical Gaussian ensembles.

One of the original motivations for the expression (\ref{Psb}) was to have a unique formula for $P(s)$ that would provide correct results in particular cases of conventional random matrix ensembles. Specifically, the function $f(\beta)$ is constructed in such a way that for the values $\beta=1,\,2,\,4$, corresponding to the Gaussian ensembles of random matrices of a given symmetry (orthogonal, unitary and symplectic, respectively) it be close to the expressions for $P(s)$ obtained in the RMT for those ensembles \cite{brody81}. As shown in Ref.~\cite{I93}, for these values of $\beta$ the dependence (\ref{Psb}) is more accurate (in a whole range of $s$) than the WD surmise typically used in the literature.

Since for $\beta=0$ Eq.~(\ref{Psb}) coincides with the Poisson distribution, and for $\beta=\infty$ it reproduces the delta-function, this interpolation formula is a perfect candidate for the description of the intermediate statistics in the closed Anderson model (\ref{Htb}). The expression similar to Eq.(\ref{Psb}) \cite{I88a} has been already used to describe $P(s)$ in finite one and two-dimensional Anderson models \cite{WM90,SS91}, resulting in the values $\beta$ from $0.05$ to $19.2$ depending on the disorder strength.

It should be stressed that in application to our model the parameter $\beta$ in Eq. (\ref{Psb}) should be considered as the parameter which {\it globally} describes the  distribution $P(s)$, rather than the parameter determining the repulsion of energy levels for very small spacings $s \ll 1$. To date, there are indications that the repulsion between neighboring energy levels in the finite Anderson model cannot be rigorously associated with the symmetry of the matrix $H_{mn}$, see Eq. (\ref{Htb}), as it happens in the case of full random matrices. Indeed, in Refs.~\cite{MS89,GRC90} it was analytically shown that for  $3\times 3$ real symmetric matrices with matrix elements $a_{13}=a_{31}=0$ and other elements random, the repulsion for small $s \ll 1$ is non-linear, $P(s) \sim s \ln (1/s)$. For $4 \times 4$ tri-diagonal random matrices the repulsion appears to be $P(s) \sim s \ln^2 (1/s)$ \cite{GRC90}. An extensive numerical study performed in Ref.~\cite{GRC90}, allows one to predict the general dependence, $P(s) \sim s \ln^{N-2} (1/s)$, for any size $N$ of tri-diagonal matrices with all non-zero elements random. For a slightly different type of random tri-diagonal matrices, the rigorous result obtained in Ref.~\cite{VV97} gives $P(s) \sim s \ln^{N-3} (1/s)$. All these results indicate that the level repulsion (defined in the limit $s \rightarrow 0$) in disordered tri-diagonal matrices strongly depends on specific properties of disorder.

\begin{figure}[h!]
\centerline{\includegraphics[width=8.0cm,angle=0]{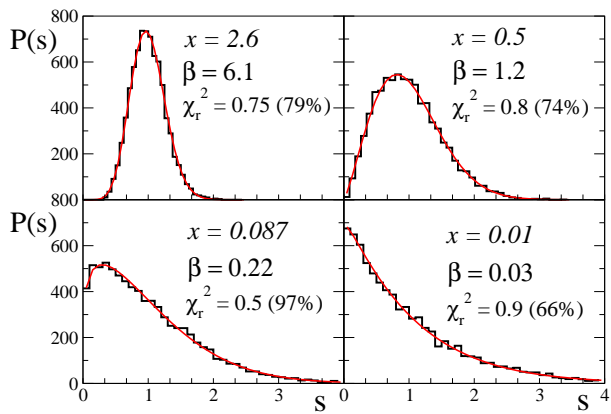}}
\caption{(Color online) Examples of $P(s)$ for $|E|<0.2$ (excluding the energies very close to 0), with $N=1000$. The data are obtained for $120$ disorder realizations with the $\chi^2$-fit to Eq.~(\ref{Psb}) for $r=40$ bins. The reduced $\chi^{2}$ statistic is shown with the corresponding confidence levels given in parenthesis.}
\label{Ps40}
\end{figure}

With Eq. (\ref{Psb}) we performed an extensive numerical study of $P(s)$ by changing the degree of disorder for a closed chain in a large range of the control parameter $x=l_\infty/N$ with $l_\infty$ defined by Eq.~(\ref{loc}), see examples in Fig.~\ref{Ps40}.

%%%%%%%===============================================%%%%%%%%%%%%%%
%%%%%%%===============================================%%%%%%%%%%%%%%

Our data demonstrate that Eq.~(\ref{Psb}) gives an amazingly good correspondence (supported by the $\chi^2$ criteria) with the numerical data in a very large range of $x$.  Therefore, one can make an unexpected conclusion that the distribution $P(s)$ for the Anderson model on a finite scale $N$ can be described by the Dyson Coulomb gas model where $s$ is the distance between the nearest particles on a ring.

Another result is that for the specific degree of disorder, namely, for $x=l_\infty/N \approx 0.435$ (see Eq. (\ref{beta-vs-loc}) below), the distribution $P(s)$ is non-distinguishable from the WD-distribution. This is in correspondence with the numerical results earlier obtained in Ref. \cite{WM90}. It is known that in the RMT an emergence of the WD-distribution is directly related to the random character of eigenstates \cite{brody81}. The numerical data \cite{WM90}, indeed, show that, in the situation when $P(s)$ is close to the WD-distribution, some of the global statistical characteristics of eigenstates are in a good correspondence with the assumption of their randomness. However, more detailed studies of eigenstates are needed in order to claim that they have the same degree of randomness as in the conventional ensembles of random matrices. Our preliminary results show that in spite of a very good correspondence of $P(s)$ to the WD-distribution, the more tiny characteristics of eigenstates demonstrate small but clear deviations from the predictions of the RMT.

%%%%%%%%%%%%%%%%%%%%%%%%%%%%%%%%%%%%%%%%%%%%%%%%%%%%%%%%%%%%%%%%%%%%%%%%%%%%%%%%%%%%%%%%%%%%%%%%%%%%%%%%%%%%%%%%%%%%

\section{Localization length vs. repulsion} 
\begin{figure}[h!]
\centerline{\includegraphics[width=6.0cm,angle=0]{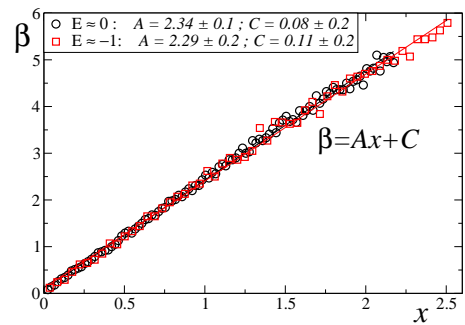}}
\caption{(Color online) Repulsion parameter $\beta$ versus $x=l_\infty/N$ for $E \approx 0$, (circles), and $E\approx -1$ (squares), see Fig.~\ref{Ps40}.}
\label{betax}
\end{figure}

Now we can establish the relation between $x$ and $\beta$. It is qualitatively clear that they express the same phenomenon of gradual transformation of standing waves into localized states but the exact relation between them was unknown. All the data are summarized in Fig. \ref{betax}. We see a precise {\it linear dependence} between $x$ and $\beta$ in a whole range of $x$ values independently of the chosen energy range. The fit of the data as $\beta=Ax+C$ gives the slope $A=2.3 \pm 0.1$ with $C$ essentially zero. This result, obtained carefully with the use of $\chi^2$ statistical criteria, shows that the repulsion parameter $\beta$ is just the normalized localization length $l_\infty$,
\begin{equation}
\label{beta-vs-loc}
\beta\approx 2.3\, \frac{l_\infty}{N}.
\end{equation}
The factor $2.3$ in Eq.~(\ref{beta-vs-loc}) can be attributed to the fluctuations of components of eigenstates.

%%%%%%%===============================================%%%%%%%%%%%%%%
%%%%%%%===============================================%%%%%%%%%%%%%%

The localization length $l_\infty$ of a given state with the site components $\psi_{n}$ can be defined through the Shannon entropy,
\beq
S=-\sum_{n=1}^{N}w_n \ln w_n
\label{shannon}
\eeq
with $w_n=\psi_n^2$. For random states with the Gaussian distribution of $\psi_n$ one gets $S=\ln (N/2.07)$. We define the normalized {\it entropic localization length}, $d_N=2.07 \exp\bra S \ket$, where $\bra...\ket$ represent an ensemble average. With this definition we have $d_N=N$ for fully chaotic eigenfunctions occupying $N$ sites. The data show that the onset of strong chaos occurs when $d_N \approx 2.1 l_\infty$ \cite{S10} which is equivalent to $\beta \approx d_N/N \approx 1$. Therefore, one arrives at the same result through an analysis of the spectrum as through the eigenstates.

According to this important result, the repulsion parameter $\beta$ is nothing but the properly rescaled localization length. This was observed in numerical studies of the kicked rotor \cite{KR} and Wigner banded random matrices \cite{WBRM}. Here, this result emerges for the standard Anderson model thus indicating a
generic effect. Very recently, the linear relation between the repulsion parameter $\beta$ and localization length was found in an experimental study of disordered elastic rods \cite{flores}.

%%%%%%%%%%%%%%%%%%%%%%%%%%%%%%%%%%%%%%%%%%%%%%%%%%%%%%%%%%%%%%%%%%%%%%%%%%%%%%%%%%%%%%%%%%%%%%%%%%%%%%%%%%%%%%%%%%%%%

\section{Open model: Non-Hermitian Hamiltonian} The scattering properties
of {\it open} systems can be formulated \cite{MW69} with the effective non-Hermitian Hamiltonian \cite{SZ92,VZ05,celardo10,S10},
\begin{equation}
{\cal H}_{mn}(E)=H_{mn}+{\cal D}(E)(\gamma^L\delta_{n,1}\delta_{m,1}+
\gamma^R \delta_{n,N}\delta_{m,N}).
\label{exact}
\end{equation}
where
\beq
{\cal D}(E)=\frac{E}{2}-\frac{i}{2}\sqrt{4-E^2},
\label{D}
\eeq
This expression is valid for any disorder $\epsilon_n$, continuum coupling $\gamma^L, \gamma^R$, and energy $E$. Near the center of the band (the general case, $-2 < E < 2$, is studied in Ref.~\cite{S10}) Eq. (\ref{exact}) reduces to the canonical form,
\begin{equation}
\label{Hnon}
{\cal H}_{mn}(E)=H_{mn}-\frac{i}{2}W_{mn}
\end{equation}
with
\beq
W_{mn}=2\pi \sum_{c=L,R} A_m^c(0) A_n^c(0).
\label{W}
\eeq
where $W_{mn}(E)$ is defined by the coupling amplitudes,
\begin{equation}
%A_n^{L,R}(E)= ( \gamma^{L,R}/\pi )^{1/2}[1-E^2/4]^{1/4}(\delta_{n,1}+\delta_{n,N}).
%A_n^{L;R}(E)= ( \gamma^{L;R}/\pi )^{1/2}[1-E^2/4]^{1/4}(\delta_{n,1;N}).
A_n^{L,R}(E)= \sqrt{ \gamma^{L,R}/\pi} \left[1-\frac{E^2}{4}\right]^{1/4}(\delta_{n,1}^{(L)}+\delta_{n,N}^{(R)}).
\label{AL}
\end{equation}
Here $H_{mn}$ in Eq.~(\ref{Htb}) describes the internal dynamics, while the non-Hermitian part $W(E)$ is factorized in terms of the coupling amplitudes $A_n^c(E)$ between the internal states $|n\ket$ and open decay channels, $c=L,R$, where $L$ and $R$ stay for left and right, respectively.

The non-Hermitian Hamiltonian (\ref{Hnon}) allows us to construct the scattering matrix $S$ in the space of channels,
\begin{equation}
S=\frac{1-i\pi K}{1+i\pi K},
\label{SR}
\end{equation}
where the reaction matrix $K$ is defined as
\begin{equation}
K^{ab}(E)=\sum_j\frac{\tilde{A}_j^a\tilde{A}_j^b}{E-E_j};
~~~~\tilde{A}_j^c=\sum_m A_m^c\psi_m^{(j)},
\label{Keigen}
\end{equation}
and $\psi_m^{(j)}$ is the $m$-component of the eigenstate $|j\rangle$ of the closed Hermitian Hamiltonian (\ref{Htb}).

For weak disorder, $\sigma^2\ll 1$, the strength of coupling to the leads is characterized by the coupling parameter,
\beq
\kappa^c=\frac{2\pi\gamma^c}{ND}.
\label{kappa}
\eeq
Here $D$ is the mean level spacing at the center of the energy band in a closed chain. By this definition, the channel transmission coefficient is maximal for {\it perfect coupling} when $\kappa^c=1$.  Below we consider the symmetric coupling, $\gamma^c\equiv \gamma$, $\kappa^c \equiv \kappa$. In all numerical simulations we take $N=1000$ sites and combine an ensemble average over a large number of realizations with a spectral average over $1000$ energies across an energy window $|E|<0.2$.

%%%%%%%%%%%%%%%%%%%%%%%%%%%%%%%%%%%%%%%%%%%%%%%%%%%%%%%%%%%%%%%%%%%%%%%%%%%%%%%%%%%%%%%%%%%%%%%%%%%%%%%%%%%%%%%%%%

\section{Transport characteristics}

For {\sl continuous} weak random potentials the problem of scattering through finite one-dimensional samples is rigorously solved by various analytical approaches (see, for example, \cite{LGP88,IKM12}). It was shown that the distribution function for the transmission coefficient $T$ (conductance) depends on the ratio of the localization length $l_\infty$ to the length $N$ of a sample. Therefore, the knowledge of the localization length (defined in the limit $N \rightarrow \infty $) gives a full description of scattering (for perfect coupling with the leads).

\begin{figure}[h!]
\centerline{\includegraphics[width=7.0cm,angle=0]{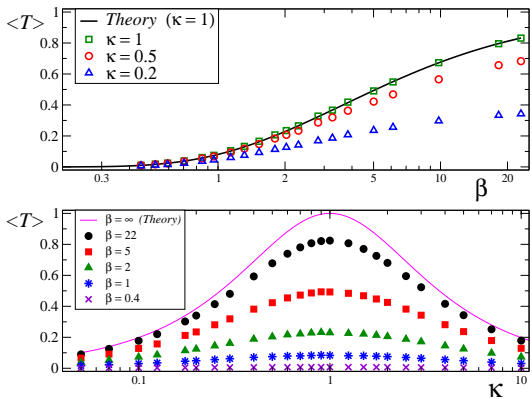}}
\caption{(Color online) Upper: $\bra T \ket$ versus $\beta = 2.3x$ for different coupling strengths $\kappa$; the smooth curve shows the analytical expression Eqs. (\ref{T-q} -\ref{T-1-2}). Lower: dependence $\bra T \ket$ versus $\kappa$ for different values of $\beta$. For $\beta \to \infty$, the average is found to tend towards $2\kappa/(1+\kappa^2)$, depicted as the smooth curve.}
\label{TTT}
\end{figure}

In contrast with continuous models, for the {\sl tight-binding model} (\ref{Htb}) there are no rigorous results, neither for the distribution of $T$ nor for the moments $\bra T^q\ket$, that would be valid everywhere inside the energy spectra, $|E|<2$, even for weak disorder. The reason is an existence of the so-called resonances for specific energies, $E_r=\pm 2\cos ( \pi r/s) $ with $r,s$ integer, for which standard perturbation theory fails. The most studied case is the band center where the Thouless expression ({\ref{loc}}) has to be corrected due to an anomaly precisely at $E = 0$ for which $l_{\infty} \simeq 105.2 / W^2$ instead of $96 / W^2$ (for details and references, see \cite{IKM12} and \cite{KY11}). On the other hand, one can expect that Eqs. (\ref{T-q}) and (\ref{T-1-2}) can be also used if the energy is not too close to the band center or band edges, with $l_\infty$ defined by Eq.~(\ref{loc}).  We have checked this conjecture for two lowest moments of the transmission coefficient, defined as $T = |S^{LR}| ^2$ \cite{LGP88,IKM12},
\begin{equation}
\bra T^q\ket=\sqrt{\frac{2x^3}{\pi}}%x^{3/2}
\exp{\left( -\frac{1}{2x}\right)}
\int_{0}^{\infty}f_{q}(z)\exp{\left( -\frac{z^2x}{2}\right)}dz ,
\label{T-q}
\end{equation}
where $x=l_\infty/N$. Here $\bra ... \ket$ stands for the ensemble average, and the functions $f_q$ for $q=1, 2$ are
\begin{equation}
f_1(z)=\frac{z^2}{\cosh{z}};~~~~~f_2(z)=\frac{2z^2+z\sinh{2z}}{4\cosh^3{z}}.
\label{T-1-2}
\end{equation}
Our numerical data, indeed, manifest the correspondence with the theoretical expression for the mean transmission with $\kappa=1$, see upper panel of Fig. \ref{TTT}. Here and in the following we express the degree of localization in terms of $\beta$ using Eq. (\ref{beta-vs-loc}).

%%%%%%%===============================================%%%%%%%%%%%%%%
%%%%%%%===============================================%%%%%%%%%%%%%%

In this figure (upper panel) we also present the transmission coefficient $\bra T \ket$ for the case of non-perfect coupling, for which the analytical results are absent. The maximal value of $\bra T \ket$ occurs for perfect coupling, $\kappa=1$. As for non-perfect coupling, the lower panel of Fig. \ref{TTT} shows that the transmission coefficient is symmetric with respect to the change $\kappa \rightarrow 1/\kappa$, that is known for the models described by full random matrices in place of $H_{mn}$ in Eq. (\ref{Hnon}) (see, e.g. \cite{SZ89}). However, our results manifest that even in the presence of strongly localized, $ \beta \ll 1 $ , or regular extended eigenstates, $\beta \gg 1 $, in a closed model, this symmetry is preserved when the system is open.

%%%%%%%===============================================%%%%%%%%%%%%%%
\begin{figure}[h!]
\centerline{\includegraphics[width=7.0cm,angle=0]{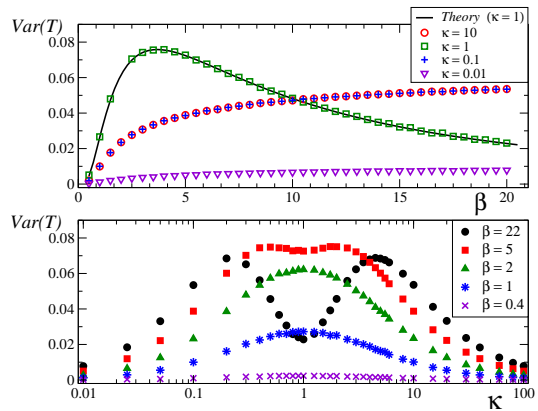}}
\caption{(Color online) Upper: $\mbox{Var}(T)$ versus $\beta$ for different values of $\kappa$. The solid curve corresponds to the analytical value for $\mbox{Var}(T)$ obtained from Eqs. (\ref{T-q} -\ref{T-1-2}). Lower: $\mbox{Var}(T)$ versus $\kappa$ for different values of $\beta$.}
\label{VarT}
\end{figure}
%%%%%%%===============================================%%%%%%%%%%%%%%

A very good correspondence was also observed for $ \bra T^2 \ket$. This gives a possibility for a discussion of the variance of the transmission, $\mbox{Var}(T)\equiv \bra T^2 \ket-\bra T \ket^2 $, characterizing {\it mesoscopic fluctuations} of the transmission. In the upper panel of Fig. \ref{VarT}, we plot $\mbox{Var}(T)$ versus $\beta$ for different fixed values of $\kappa$. Once more, an excellent agreement between the analytical expressions given by Eqs. (\ref{T-q} -\ref{T-1-2}) and the numerical data based on the discrete model (\ref{Hnon}) can be seen when $\kappa=1$. For this case, the variance reaches a maximum for $\beta \approx 4.0$ which is much larger than that corresponding to maximal internal chaos ($\beta =1)$, as one may naively expect. For $\beta \approx 4$, the eigenstates are extended; however, they are neither strongly chaotic nor regular. Our data demonstrate a non-monotonic dependence of $\mbox{Var}(T)$ on both sides of $\kappa =1$ for some critical value $\beta_{cr} \sim 4$ and above. Below this critical value the single maximum of $\mbox{Var}(T)$ occurs at perfect coupling whereas above $\beta_{cr}$ there are two maxima on both sides.

Moving away from non-perfect coupling, a clear symmetry is seen for the mutually inverse values $\kappa =10$ and $0.1$; this symmetry is also observed in the lower panel of the same figure where we plot the variance of $T$ versus $\kappa$ for different values of $\beta$. Such a symmetry was repeatedly mentioned in the literature, see for example \cite{our}; it follows from physics of many intrinsic states coupled to the same decay channels; one of the first derivations follows from the Moldauer-Simonius relation \cite{MS}. With $\kappa$ growing beyond the super-radiant transition at $\kappa\sim 1$, the broad state becomes a part of the background while the narrow resonances return to the non-overlap regime.

\begin{figure}[h!]
\centerline{\includegraphics[width=6cm,angle=0]{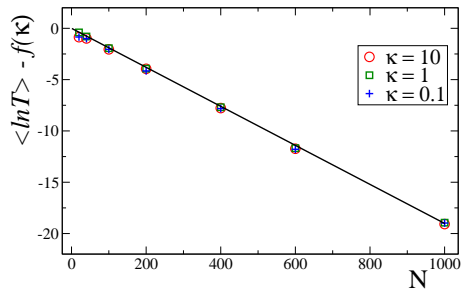}}
\caption{(Color online) $\bra \ln{T}\ket - f(\kappa)$ vs $N$ for different values of $\kappa$. Here $f(\kappa) = 2\ln{\left[4\kappa/(1+\kappa)^2\right]}$, see Eq. (\ref{lnTf}), and $W=1; ~l_{\infty}=105.2$ for $E=0$.}
\label{lnT}
\end{figure}

The analytical estimates (\ref{T-q}) and (\ref{T-1-2}) for $\bra T \ket$ are valid for perfect coupling, $\kappa=1$, only. On the other hand, one can  obtain an analytical expression for $\langle \mbox{ln}~T \rangle$ which is valid for {\it any} value of $\kappa$ and again reveals the symmetry $\kappa\Leftrightarrow 1/\kappa$,
\begin{equation}
\bra \ln{T} \ket = -\frac{2N}{l_{\infty}}+2\ln\left[ \frac{4\kappa}{(1+\kappa)^2}\right].
\label{lnTf}
\end{equation}
It can be derived via the product of $N+2$ transfer matrices, by tracing out two (non-random) matrices describing the coupling to the leads. This expression can be used for the definition of the localization length $l_\infty$ in the presence of non-perfect coupling. Our numerical data for $\bra \ln{T}\ket$  manifest excellent correspondence with Eq.~(\ref{lnTf}), at the band center, see Fig. \ref{lnT}, as well as for different values of $\beta$ (not shown).

%%%%%%%===============================================%%%%%%%%%%%%%%
%%%%%%%===============================================%%%%%%%%%%%%%%

%%%%%%%%%%%%%%%%%%%%%%%%%%%%%%%%%%%%%%%%%%%%%%%%%%%%%%%%%%%%%%%%%%%%%%%%%%%%%%%%%%%%%%%%%%%%%%%%%%%%%%%%%%%%%%%%

\section{Correlations}

The knowledge of the scattering matrix $S$ defined by Eq. (\ref{SR}), allows one to study many details of scattering. One of the problems of both theoretical and experimental interest is to understand how the correlations between different cross sections depend on the degree of localization for different coupling strengths. This problem was recently addressed in Refs.~\cite{our,SISZB08,SICZB09} where  the non-Hermitian Hamiltonians
contained, apart from a diagonal part, band-like random symmetric matrices. Such models are typically studied in application to open complex systems, like heavy nuclei and quantum dots. There, the amplitudes $A_m^{c}$ connecting intrinsic states $|m\rangle$ with many open channels, $c=1,...,M \gg 1$, were assumed to be completely random and independent of internal dynamics. In contrast, in our case the derivation of the Hamiltonian (\ref{Hnon}), for specified internal disorder $\{\epsilon_n\}$, does not involve any additional assumption of randomness for the coupling to continuum. Our main interest  here is to understand how the properties of cross section correlations depend on whether the internal eigenstates are localized or extended, and how  the correlations depend on the coupling to continuum.

The average cross section can be divided into two parts corresponding to {\it direct} and {\it fluctuating} processes, respectively (see, e.g. \cite{our}).  The direct processes correspond to a very fast passage of an incoming particle (or wave) through the scattering region, thus resulting in broad short-lived states of an open system. In contrast, the fluctuating parts of cross sections describe narrow long-lived states (resonances) known in nuclear physics as ``compound" states. In our model with $M=2$ channels, the transmission and reflection coefficients are defined as follows,
\begin{equation}
T=|S^{LR}|^2=|S^{LR}_{fl}|^2\, ; ~~~ R=|S^{LL}|^2=|\langle S^{LL}\rangle + S^{LL}_{fl}|^2,
\label{T-R}
\end{equation}
where $``fl"$ stands for fluctuating parts. With the usual definition of the (partial) cross section, $\sigma^{ab}=|\delta_{ab}-S^{ab}|^2$, we define the fluctuating cross sections as $\sigma^{LR}=|S_{fl}^{LR}|^2 $ and $\sigma^{LL}=|S_{fl}^{LL}|^2$. Therefore, the average fluctuating part of the two cross sections can be expressed in terms of the average transmission, $T$, and reflection, $R$, coefficients:
\begin{equation}
\langle\sigma^{LR}\rangle=\langle T \rangle; ~~~~~~ \langle\sigma^{LL}\rangle=\langle R \rangle-\langle S^{LL}\rangle ^2.
\label{mean}
\end{equation}

The average elements of the scattering matrix can be written as [compare with Eq. (13)]
\begin{equation}
\langle S^{ab}\rangle = \frac{1-\kappa}{1+\kappa}\delta_{ab},
\label{meanS}
\end{equation}
This expression is well known in the RMT (see, for example, \cite{SZ89}). One can show that it is also valid for the considered Anderson model \cite{S10}. Note that for perfect coupling, $\kappa=1$, the mean values $\langle\sigma^{LL}\rangle$ and $\langle\sigma^{LR}\rangle$ of the cross sections are nothing but the average transmission and reflection coefficients, respectively.

Two types of correlations of our interest are defined as
\begin{eqnarray}
\label{coeff}
C_1 & = & \langle \sigma^{LL} \sigma^{LR} \rangle - \langle\sigma^{LL}\rangle \langle \sigma^{LR}\rangle \ , \\
C_2 & = & \langle \sigma^{LL} \sigma^{RR} \rangle - \langle\sigma^{LL}\rangle \langle \sigma^{RR}\rangle \ .
\label{sym2}
\end{eqnarray}

%%%%%%%===============================================%%%%%%%%%%%%%%
\begin{figure}[h!]
\centerline{\includegraphics[width=6.5cm]{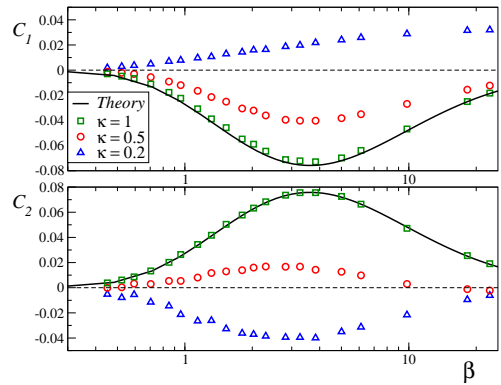}}
\caption{(Color online) Correlations $C_1$ (upper) and $C_2$ (lower) versus $\beta$ for $\kappa =1, ~0.5, ~0.2$. The solid curves are the analytical expressions (\ref{T-q}-\ref{T-1-2}) for $-{\mbox Var}(T)$ and ${\mbox Var}(T)$ shown in upper and lower panels, respectively.}
\label{VarT-xi}
\end{figure}
%%%%%%%===============================================%%%%%%%%%%%%%%

\begin{figure}[h!]
\centerline{\includegraphics[width=6.5cm]{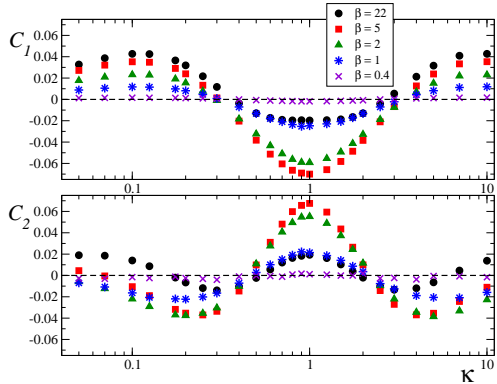}}
\caption{(Color online) Correlations $C_1$ (upper) and $C_2$ (lower) versus $\kappa$ for different values of $\beta$ as indicated.}
\label{VarT-k}
\end{figure}

In literature they are referred to as the {\it covariances}, widely studied in connection with transmission through waveguides with bulk and surface scattering (see, for example, \cite{GSNS02} and references therein). The result of computation of the correlations (\ref{coeff}-\ref{sym2}) is presented in Fig. \ref{VarT-xi} as a function of the repulsion parameter $\beta$ for different values of $\kappa$. For perfect coupling, the data are well described by the analytical expressions, since in this case  we simply have
\beq
C_1=-\mbox{Var}(T), \qquad C_2=\mbox{Var}(T).
\label{CC}
\eeq
Referring to Fig. \ref{VarT-k} for non-perfect coupling, one has to stress the following. First, both for $C_1$ and $C_2$ there is a critical value of coupling, for which the sign of correlation changes for any value of $\beta$. Second, the symmetry $\kappa \Leftrightarrow 1/\kappa$  is seen for any value of the repulsion parameter $\beta$, therefore, for any degree of localization in the closed model. As a whole, the dependence of the correlations $C_1$ and $C_2$ on the coupling $\kappa$ and on the degree of internal chaos (in our case defined by the repulsion parameter $\beta$), qualitatively agrees with that found in Refs.\cite{SICZB09}  for random matrix models.

%%%%%%%===============================================%%%%%%%%%%%%%
%%%%%%%===============================================%%%%%%%%%%%%%

%%%%%%%%%%%%%%%%%%%%%%%%%%%%%%%%%%%%%%%%%%%%%%%%%%%%%%%%%%%%%%%%%%%%%%%%%%%%%%%%%%%%%%%%%%%%%%%%%%%%%%%%%%%%%%%%%

\section{Conclusion} We studied the transport properties of the 1D Anderson model in dependence on the degree of internal chaos and strength of coupling to continuum. We found that the level spacing distribution $P(s)$ for a closed model of finite size $N$ is well described by the phenomenological expression (\ref{Psb}) where the repulsion parameter $\beta$ changes from $\beta=0$ to $\beta =\infty$. This expression is originated from the two-dimensional Coulomb gas model with the temperature $1/\beta$, and gives the distribution of spacings between nearest charged particles moving on a ring. This fact may be used for further analytical studies of the spectrum statistics in the finite Anderson model.

In the closed model we established an important linear relation between the parameter $\beta$ of spectral statistics and the normalized localization length, $l_{\infty}/N$, of the  eigenfunctions. This result still awaits for a rigorous analysis. The important point is that the parameter $\beta$ can be used to describe the transformation of extended standing waves into localized states, when increasing the degree of disorder. In passing from extended to localized states, our data clearly manifest the Wigner-Dyson distribution occurring in a quite narrow region of the disorder strength, for $\beta \approx 1$.

Opening the system at the ends we used the effective non-Hermitian Hamiltonian to study the transport properties of the model. For perfect coupling we demonstrated that both the transmission coefficient (conductance) and its variance can be analytically described by the theoretical expressions developed for disordered models with continuous potentials. This fact allows one to fully predict how the mesoscopic fluctuations depend on the degree of internal chaos quantified by the spectral parameter $\beta$, or, the same, by the normalized localization length. For non-perfect couplings, we have developed the expression for the mean logarithm of conductance which is confirmed by our numerical data. Our extensive numerical study of a non-perfect coupling reveals specific properties of transport characteristics in dependence on {\it both} internal chaos and coupling strength to continuum.

Our special interest was in the study of correlations  between two cross sections related to the transmission and reflection of scattering waves. The data show that the dependence of these correlations on the degree of disorder and on the coupling strength with the continuum, is qualitatively of the same type as found in Ref.\cite{SICZB09} for the models described by random non-Hermitian matrices. Since our original model (\ref{Htb}) is adequate to one-mode waveguides with inserted scatterers \cite{kuhl}, one can suggest that the properties of scattering revealed in our study can be experimentally observed.

\section{Acknowledgements} V.Z. acknowledges the NSF grants PHY-0758099 and PHY-1068217, F.M.I. acknowledges support from CONACyT grant N-161665, and S.S acknowledges the support from the Leverhulme Trust.

%%%%%%%%%%%%%%%%%%%%%%%%%%%%%%%%%%%%%%%%%%%%%%%%%%%%%%%%%%%%%%%%%%%%%%%%%%%%

\end{document}